\documentclass{elsart}
\journal{New Astronomy}

\usepackage{natbib}

\usepackage{graphicx}
\usepackage{epsfig}

\usepackage{amssymb}

\begin{document}

\begin{frontmatter}



 \title{Tracing early structure formation with massive starburst galaxies and their implications for reionization}


\author{Kentaro Nagamine$^1$, Renyue Cen$^2$, Steven R. Furlanetto$^3$,}
\author{Lars Hernquist$^4$, Christopher Night$^4$, Jeremiah P. Ostriker$^2$,} 
\author{Masami Ouchi$^5$}
\address{$^1$University of California, San Diego, {\rm Email: knagamine@ucsd.edu}\\
$^2$Princeton University Observatory, $^3$California Institute of Technology,\\ $^4$Harvard University, $^5$Space Telescope Science Institute}


\begin{abstract}

Cosmological hydrodynamic simulations have significantly improved over the 
past several years, and we have already shown that the observed properties of 
Lyman-break galaxies (LBGs) at $z=3$ can be explained well by the massive 
galaxies in the simulations. Here we extend our study to $z=6$ and 
show that we obtain good agreement for the LBGs at the bright-end of 
the luminosity function (LF). Our simulations also suggest that the 
cosmic star formation rate density has a peak at $z= 5-6$, and that 
the current LBG surveys at $z=6$ are missing a
significant number of faint galaxies that are dimmer than the current 
magnitude limit. Together, our results suggest that the universe could be
reionized at $z=6$ by the Pop II stars in ordinary galaxies. 

We also estimate the LF of Lyman-$\alpha$ emitters (LAEs) at $z=6$ 
by relating the star formation rate in the simulation to the Ly$\alpha$ 
luminosity. 
We find that the simulated LAE LFs agree with the observed data provided 
that the {\it net} escape fraction of Ly$\alpha$ photon is 
$f_{Ly\alpha} \leq 0.1$. We investigate two possible scenarios for 
this effect: (1) all sources in the simulation are uniformly dimmer by 
a factor of 10 through attenuation, and (2) one out of ten LAEs randomly lights 
up at a given moment. We show that the correlation strength of the LAE spatial 
distribution can possibly distinguish the two scenarios.  
\end{abstract}


\end{frontmatter}


\section{Introduction}
One of the main goals of computational cosmology is to draw a complete, 
self-consistent picture of galaxy formation from first principles. 
Given the matter fluctuations in the initial conditions that are 
motivated by the inflationary theories, 
we can simulate the evolution of structure and the formation of galaxies 
using the laws of gravity and hydrodynamics under the cosmological model 
that is suggested by the observations such as the WMAP 
(Spergel et al. 2003) and Type I supernovae (e.g. Riess et al. 1998; 
Perlmutter et al. 1999).

Over the past several years, cosmological hydrodynamic simulations
have improved significantly, and the simulations are now able to reproduce 
reasonable population of galaxies whose properties we can directly 
compare to the actual observations. In a series of past publications 
(Nagamine 2002; Nagamine et al. 2004ab, 2005a; Night et al. 2005), 
we have shown that our hydrodynamic simulations can account for the 
properties of LBGs at $z=3-6$ very well if we associate them with massive 
starburst galaxies embedded in massive dark matter halos. 
Nagamine et al. (2005b) also discussed the near-IR properties of massive 
galaxies and Extremely Red Objects (EROs) in the simulations.  
In this conference proceedings, we summarize the study of LBGs at $z=6$ 
as well as the Lyman-$\alpha$ emitters (LAEs) in cosmological 
hydrodynamic simulations.


\section{Cosmological Hydrodynamic Simulations}
\label{sec:simulation}

We utilize two different types of cosmological hydrodynamic simulations
based on a cold dark matter model:
an Eulerian mesh code {\small TIGER} (Cen, Nagamine \& Ostriker 2004) 
with Total Variation Diminishing (TVD; Ryu et al. 1993) shock capturing 
scheme, and a Smoothed Particle Hydrodynamics (SPH) code {\small GADGET-2}
(Springel 2005) which employs a novel entropy-conserving formulation
(Springel \& Hernquist 2002). 
Both simulations include standard physical treatments such as
radiative cooling/heating, uniform UV background radiation, supernova
feedback, and star formation. The SPH simulation also includes
feedback by galactic wind and a multiphase ISM model for star formation
(Springel \& Hernquist 2003a,b).   
At each time step of the simulation, 
some fraction of gas is converted to star particles based on a 
Schmidt-type law in regions of very high density to model galaxy 
formation. Each star particle is tagged with physical quantities
such as stellar mass, formation time, and metallicity, which enable 
us to apply a population synthesis technique on each star particle
and obtain luminosity output. Collections of star particles are 
identified with a grouping algorithm and identified as galaxies. 
Table~1 lists important parameters of the simulations, and we 
refer the readers to Nagamine et al. (2005a,b) for more details on 
particular simulation runs that we utilize for our study.
All simulations assume a standard flat $\Lambda$ cosmology. 

\begin{table}
\begin{center}
\begin{tabular}{cccccc}
\hline
Run &  Boxsize & ${N_{\rm mesh/ptcl}}$ & $m_{\rm DM}$ & $m_{\rm gas}$ & $\epsilon$ \\
\hline
\hline
SPH Q5  & 10.   & $324^3$ &  $2.1\times 10^6$ & $3.3\times 10^5$ & 1.2 \cr
SPH Q6  & 10.   & $486^3$ &  $6.3\times 10^5$ & $9.7\times 10^4$ & 0.8 \cr
SPH D5  & 33.75 & $324^3$ &  $8.2\times 10^7$ & $1.3\times 10^7$ & 4.2 \cr
SPH G6  & 100.  & $486^3$ &  $6.3\times 10^8$ & $9.7\times 10^7$ & 5.3 \cr
\hline
TVD N864L22 & 22. & $864^3$ & $8.9\times 10^6$ & $2.2\times 10^5$  & 25.5\cr
TVD N1024L85 & 85. & $1024^3$ & $1.6\times 10^8$ & $3.6\times 10^6$ & 83.0\cr
\hline
\end{tabular}
\vspace{0.5cm}
\caption{SPH and TVD simulations employed in this study.  
The box size is in units of $h^{-1}$Mpc. 
For the SPH, the (initial) number of gas particles 
$N_{\rm P}$ is equal to the number of dark matter particles, so the 
total particle count is twice ${N_{\rm ptcl}}$. $m_{\rm DM}$ and 
$m_{\rm gas}$ are the masses of dark matter and gas particles (mean 
baryonic mass per cell for TVD) in units of $h^{-1}M_\odot$, 
respectively. $\epsilon$ is the comoving gravitational softening length 
for the SPH (cell size for TVD) in units of $h^{-1}$kpc. 
}
\label{table:sim}
\end{center}
\vspace{0.5cm}
\end{table}


\section{Galaxy LF at $z=6$ \& Cosmic Star Formation History}
\label{sec:LBG}

\begin{figure}[!t]
\centerline{\psfig{file=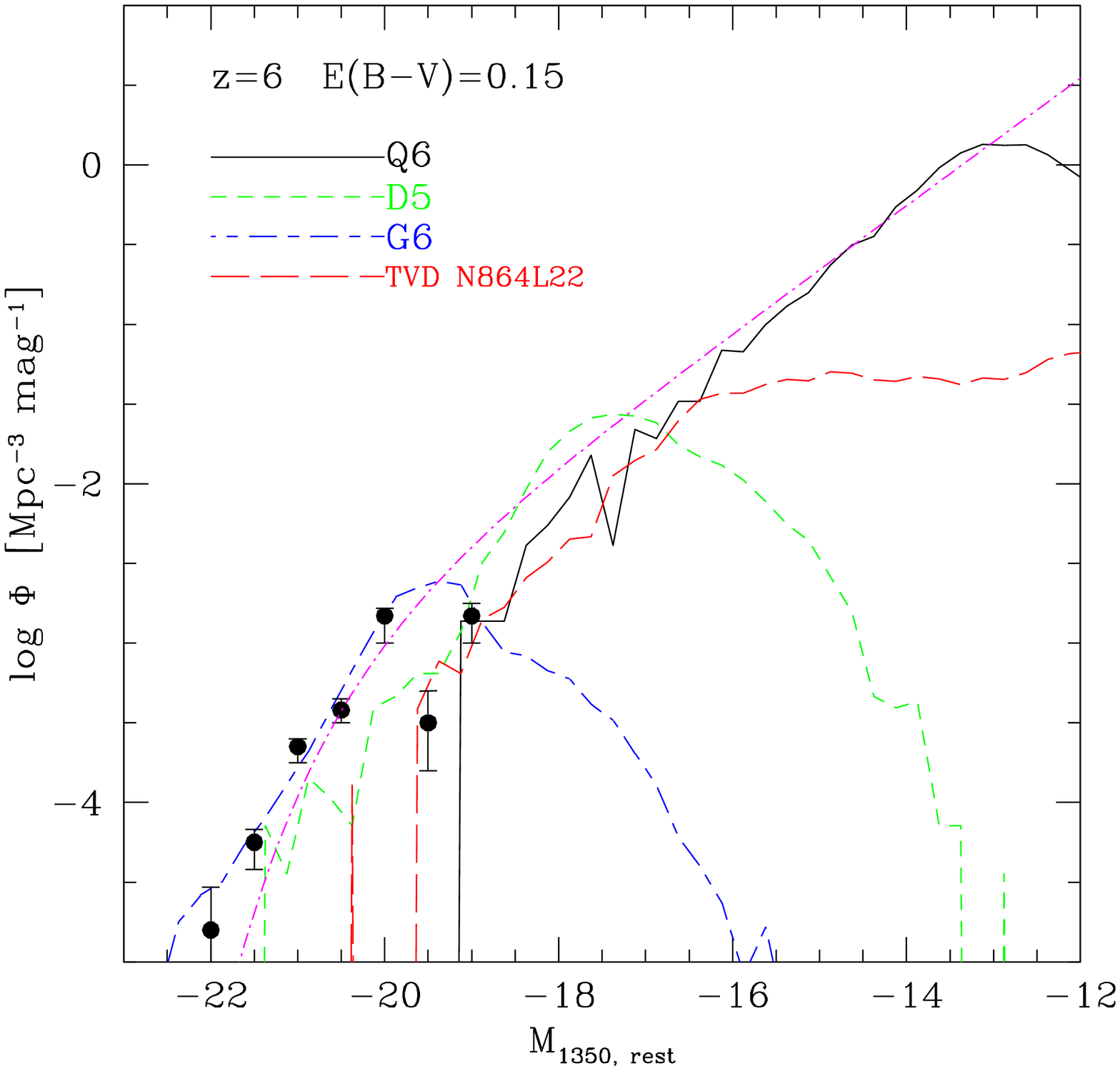,width=2.7in}\psfig{file=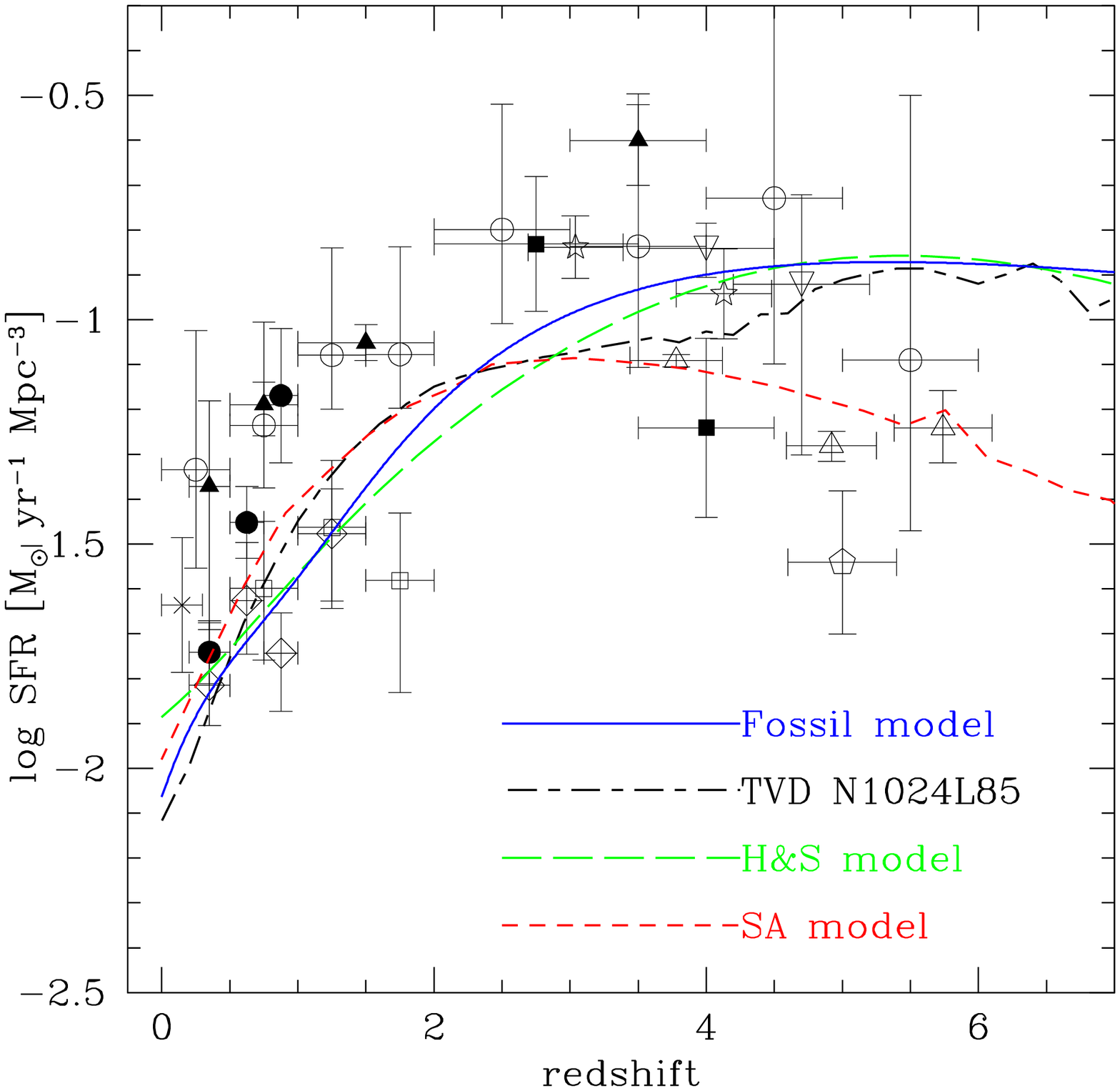,width=2.7in}}
\caption{{\bf \it Left:} Luminosity function of galaxies at $z=6$ in simulations.
Uniform extinction of $E(B-V)=0.15$ is assumed.  Data points are from 
Bouwens et al. (2004). 
{\bf \it Right:} Cosmic star formation history in hydro simulations 
and other models. Note that all models except the semianalytic model 
(SA model) have a peak of SFR density at $z\gtrsim 5$. H\&S model is
the one by Hernquist \& Springel (2003) based on the results of 
SPH simulations. For the details on the extinction corrected data points, 
see Nagamine et al. (2004b). The `Fossil' model is from Nagamine et al. 
(2005, in preparation).  
}
\label{fig:LBG}
\vspace{0.4cm}
\end{figure}

The luminosity function (LF) of simulated LBGs at $z=6$ is shown 
in the left panel of Fig.~\ref{fig:LBG}. 
It shows that the simulated LF agrees well with 
the observed one by Bouwens et al. (2004) at the bright-end. 
It is encouraging that we obtain good agreement with observations 
for the LFs at $z=3-6$ throughout, as well as with the rest-frame UV 
colors (Night et al. 2005).
This suggests that the simulated LBGs have realistic star formation 
rates and histories compared to those of the actual LBGs. A notable 
feature of simulated LFs is that the faint-end slope is quite steep with 
$\alpha \simeq -2.0$ in the rest-frame magnitude range $-19 < M_{1350} <
-16$. This magnitude range is even beyond the current magnitude limit of 
rest-frame $M_{1350}\simeq -19$.

In the right panel of Fig.~\ref{fig:LBG}, we show the cosmic star 
formation history as a function of redshift. Both hydro simulations
(Nagamine et al. 2001, 2004b) and the model by Hernquist \& Springel (2003) 
suggest that the peak of the SFR density is at $z\ge 5$, which is quite 
different from the predictions of some of the semi-analytic models (e.g. 
Kauffmann et al. 1999; Cole et al. 2000; Somerville et al. 2001; 
Menci et al. 2002). 
Together with the steep faint-end slope of the LF, our results suggest 
that there would be enough ionizing photons to reionize the Universe at 
$z=6$ with the Pop II stars in standard galaxies (Nagamine et al. 2005, 
in preparation).


\section{Large-scale scale structure at $z=6$ and Lyman-$\alpha$ Emitters}

\begin{figure}[!t]
\centerline{\psfig{file=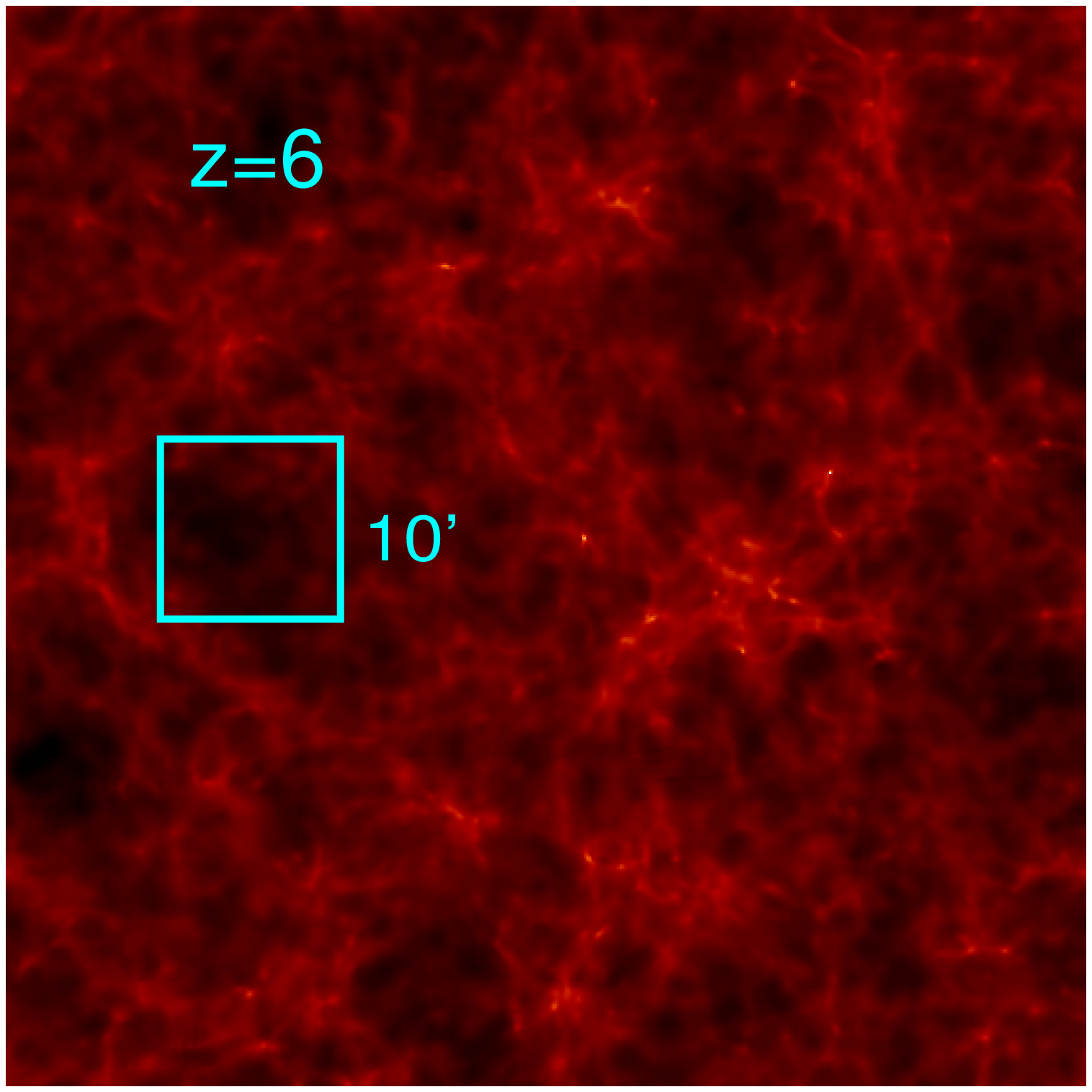,width=2.7in}
\hspace{0.01cm}
\psfig{file=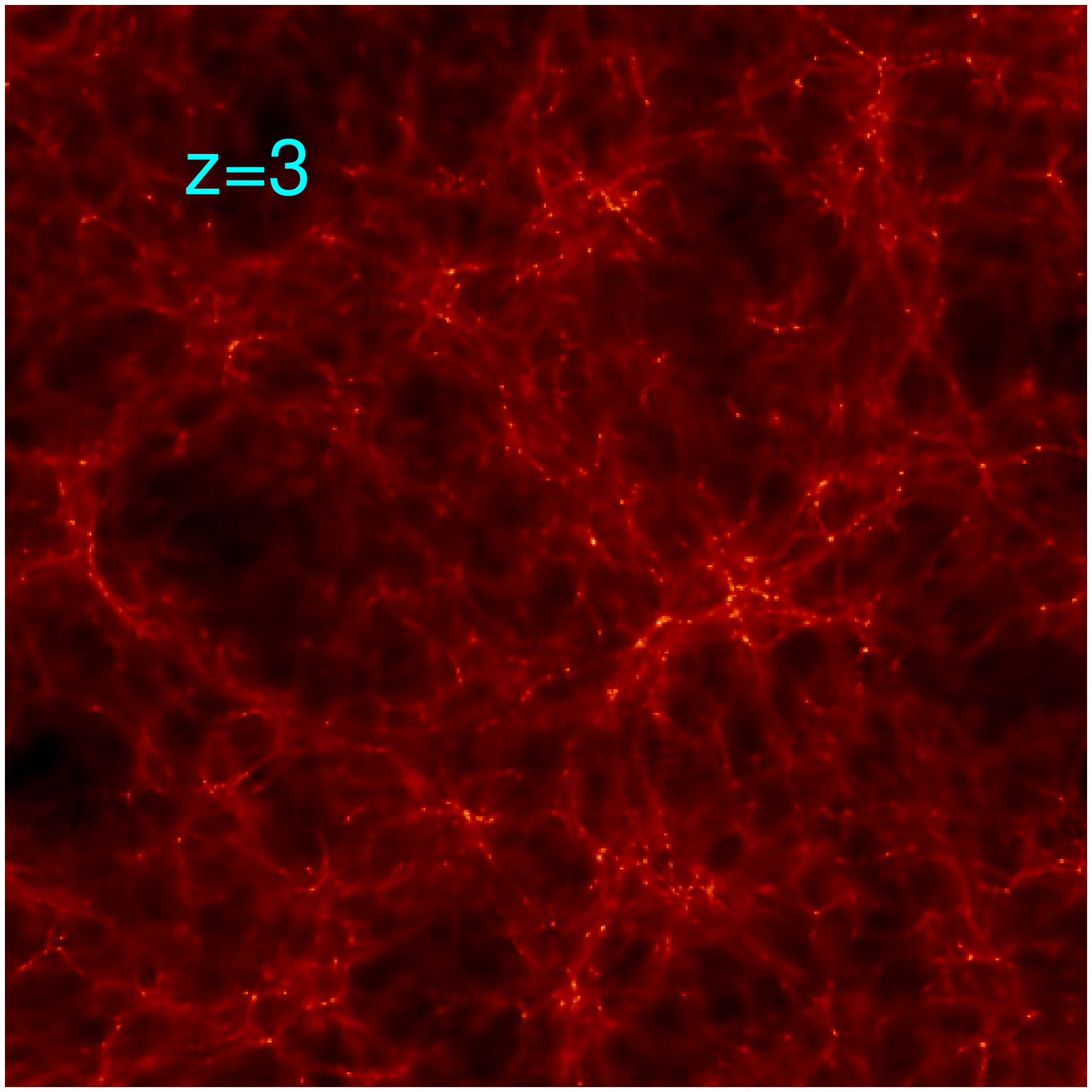,width=2.7in}}
\caption{Projected dark matter density field at $z=6$ (left) and 3 (right)
from the SPH G6 run. 
Each panel has a size of comoving 143 Mpc on a side, corresponding to 
1~degree at $z=6$. The small box in the left panel shows an example 
survey area of (10 arcmin)$^2$.
}
\label{fig:dm}
\vspace{0.5cm}
\end{figure}

Figure~\ref{fig:dm} shows the projected dark matter density field
at $z=6$ (left) and 3 (right). The left panel shows that the large-scale 
structure is already roughly in place by $z=6$ with large voids. 
Each panel has a size of comoving 143 Mpc on a side, corresponding to 
1~degree at $z=6$.
The filaments become tighter and high density peaks become more prominent 
by $z=3$. In our simulations, LBGs and LAEs are centered on high 
density peaks that appear bright in this figure. The figure shows that 
a large survey area ($\sim$ 1 degree) is needed in order to obtain a 
representative sample of LBGs or LAEs at $z=6$. If a (10 arcmin)$^2$ 
field-of-view of a survey was (unluckily) placed on a void as shown 
in this picture, the number density of the source would be significantly 
underestimated relative to the cosmic mean due to a strong cosmic variance
effect.

\section{Luminosity function of LAEs at $z=6$}

\begin{figure}[!t]
\centerline{\psfig{file=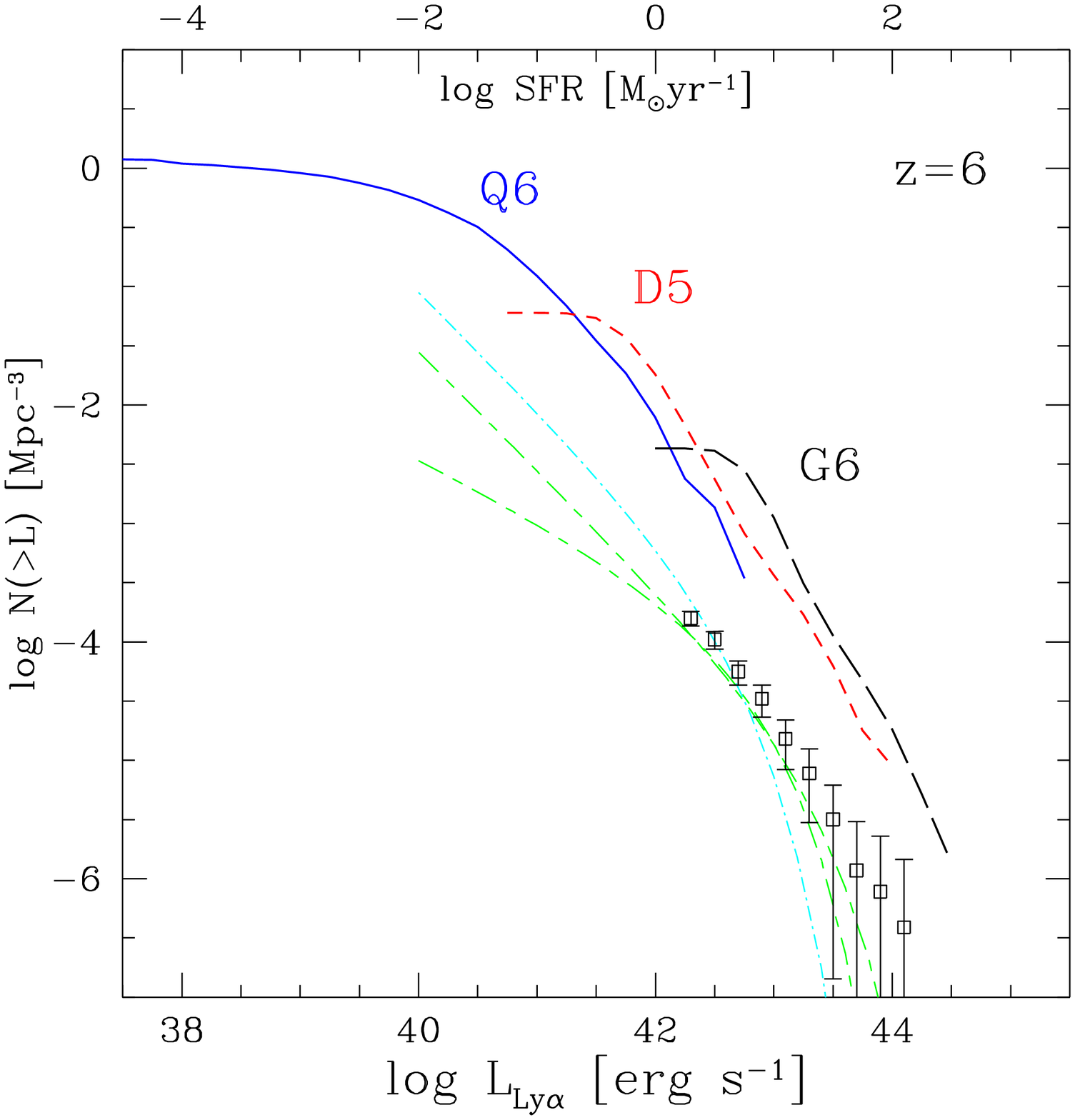,width=2.7in}
\hspace{0.01cm}
\psfig{file=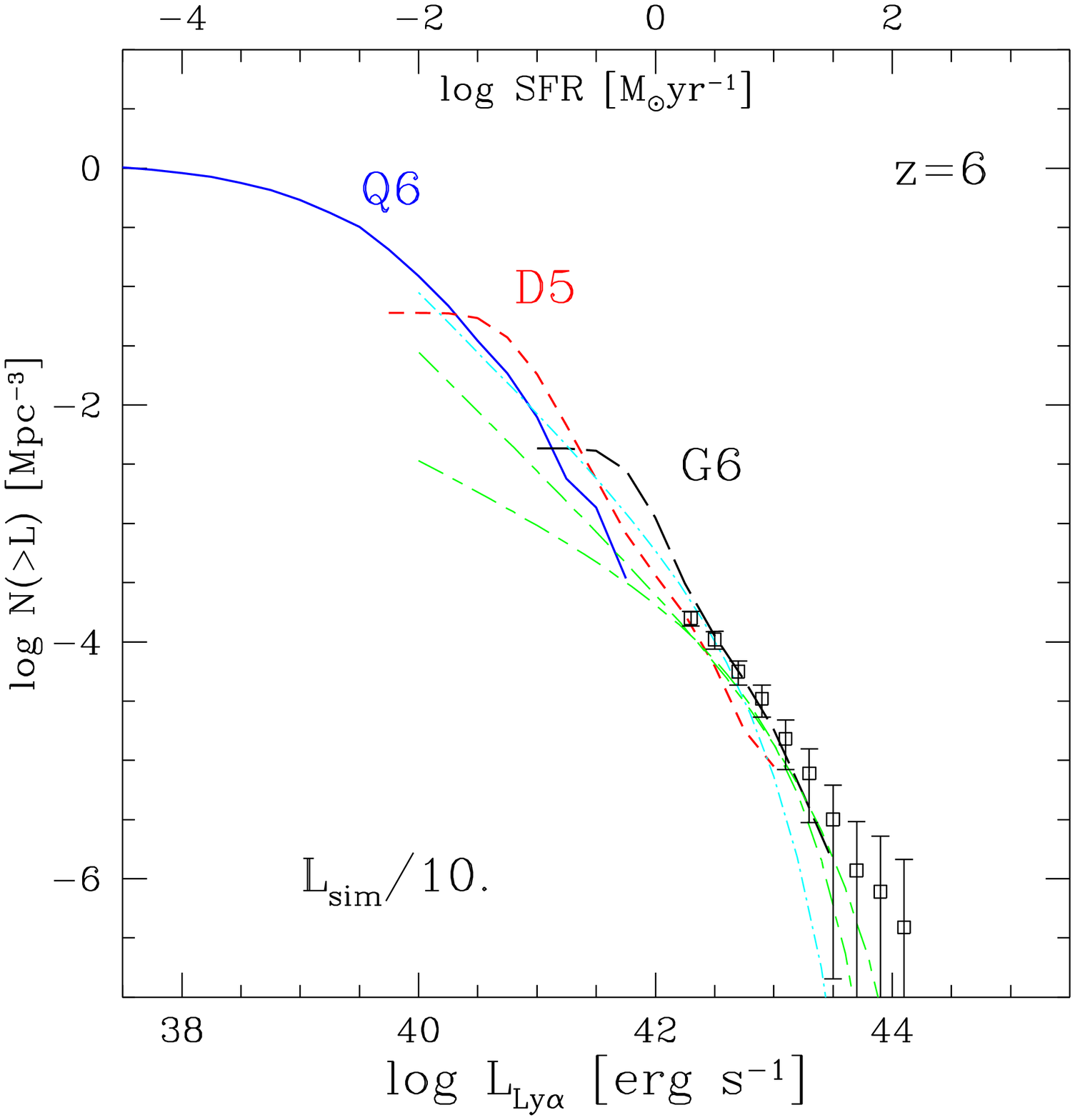,width=2.7in}}
\caption{Cumulative luminosity function of LAEs from 3 different SPH 
simulations. Left panel is directly estimated from the simulation using 
the method described in the text, and the right panel is the 
`uniform dimming' scenario where all sources are dimmed by a factor of 10. 
The larger box-size simulation covers the brighter end of the LF due 
to more numerous massive halos.
}
\label{fig:cum}
\vspace{0.5cm}
\end{figure}

We estimate the Ly$\alpha$ luminosity of each galaxy in the simulation 
from the instantaneous star formation rate using the relation 
$L_{Ly\alpha} =10^{42}$ erg s$^{-1}$ per 1\,$M_\odot$ yr$^{-1}$ (e.g. 
Leitherer et al. 1999). This conversion factor could be higher if 
the IMF is top heavy. 
In the left panel of Figure~\ref{fig:cum}, we show the cumulative 
luminosity function of LAEs at $z=6$ estimated from the simulations 
as we described above. We find that the simulations overpredict 
the LF by a factor of $\approx 10$ compared to the observational 
estimate by Ouchi et al. (2005b, in preparation) for 
the $z\sim 6$ LAE sample presented in Ouchi et al. (2005a). 
Since their field-of-view was $\sim$ 1.5 degrees, the sample is 
large enough that it shouldn't be severely affected by the cosmic variance,
and their result also agrees with that of Malhotra \& Rhoads (2004) 
within the error bars. Furlanetto et al. (2005) have also found
the same overestimate of LAE LF in simulations at lower redshifts.

There are at least two possible ways to reconcile the overprediction of 
the simulated LF relative to the observed one: (1) all the LAEs are attenuated 
by a factor of $\sim 10$ (hereafter `uniform dimming' scenario), or 
(2) one out of ten LAEs randomly `turns on' at a given time (hereafter
`random sampling' scenario). In the latter case, the reason for the 
randomness would be primarily due to the random geometry of the gas 
clouds around star-forming regions which block the Ly$\alpha$ 
photons traveling towards us, and not due to the fluctuation of 
star formation activity. The fluctuating SFR is already reflected 
in the distribution of the Ly$\alpha$ luminosity, and the simulation 
box is large enough to smooth out such fluctuations when averaged
over the entire simulation volume.  
For these reasons, here we take the 1:10 ratio at face value as 
a random sampling of the sources and see what we can learn from this 
simple exercise. In the real Universe, a combination of the above
two effects may be at play.  

The right panel of Fig.~\ref{fig:cum} shows the `uniform dimming' 
scenario where all LAEs are uniformly dimmed by a factor of 10. 
This uniform dimming can be regarded as a {\it net} result of 3 
different physical processes: First is the fact that a fraction 
$f_{\rm esc}$ of all ionizing photons escape from the galaxy and 
do not contribute to the creation of Ly$\alpha$ photons. 
Second is the dust attenuation by a factor of $e^{-\tau_{\rm dust}}$. 
Third is that only a fraction $f_{\rm IGM}$ of Ly$\alpha$ 
photons penetrates the IGM and reaches us (e.g. Barton et al. 2004), 
although this effect is only significant before reionization 
($z\gtrsim 6$). 
This can be summarized as follows: 
$F_{\rm Ly\alpha, obs} = f_{Ly\alpha}\, F_{\rm Ly\alpha, em} = 
e^{-\tau_{\rm dust}}\, (1-f_{\rm esc})\,f_{\rm IGM}\, F_{\rm Ly\alpha, em}$, 
where $f_{Ly\alpha}$ characterizes the {\it net} escape fraction 
of Ly$\alpha$ photons from LAEs. Our simulations suggest 
$f_{Ly\alpha}\sim 0.1$ for a conversion factor of 
$L_{Ly\alpha} =10^{42}$ erg s$^{-1}$ per 1\,$M_\odot$ yr$^{-1}$. 
If the IMF is top heavy and the true value of the conversion factor 
is larger than the above, then the value of $f_{Ly\alpha}$ will become 
proportionally smaller.


\section{Correlation function of LAEs}

\begin{figure}[!t]
\centerline{\psfig{file=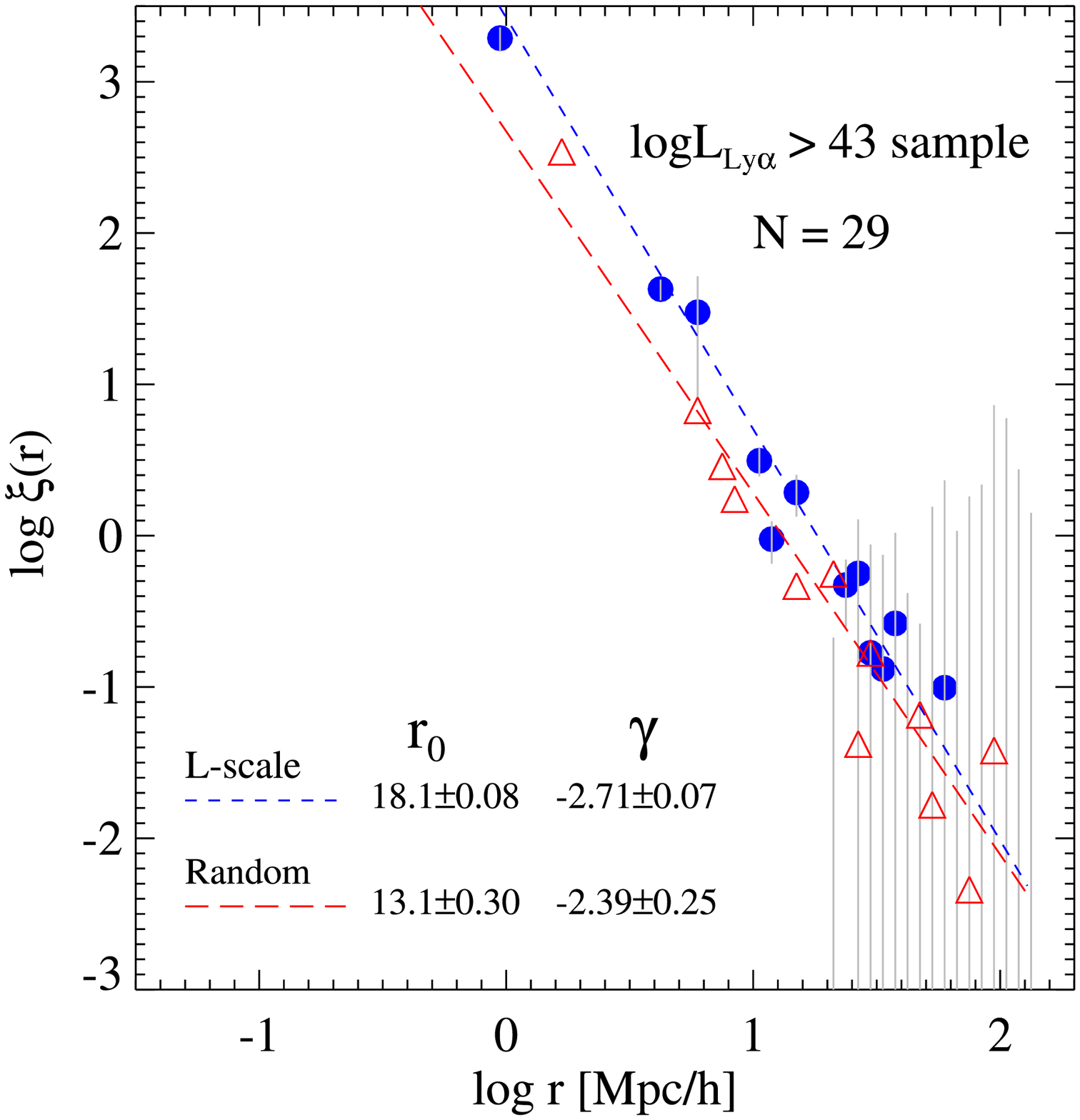,width=2.7in}\psfig{file=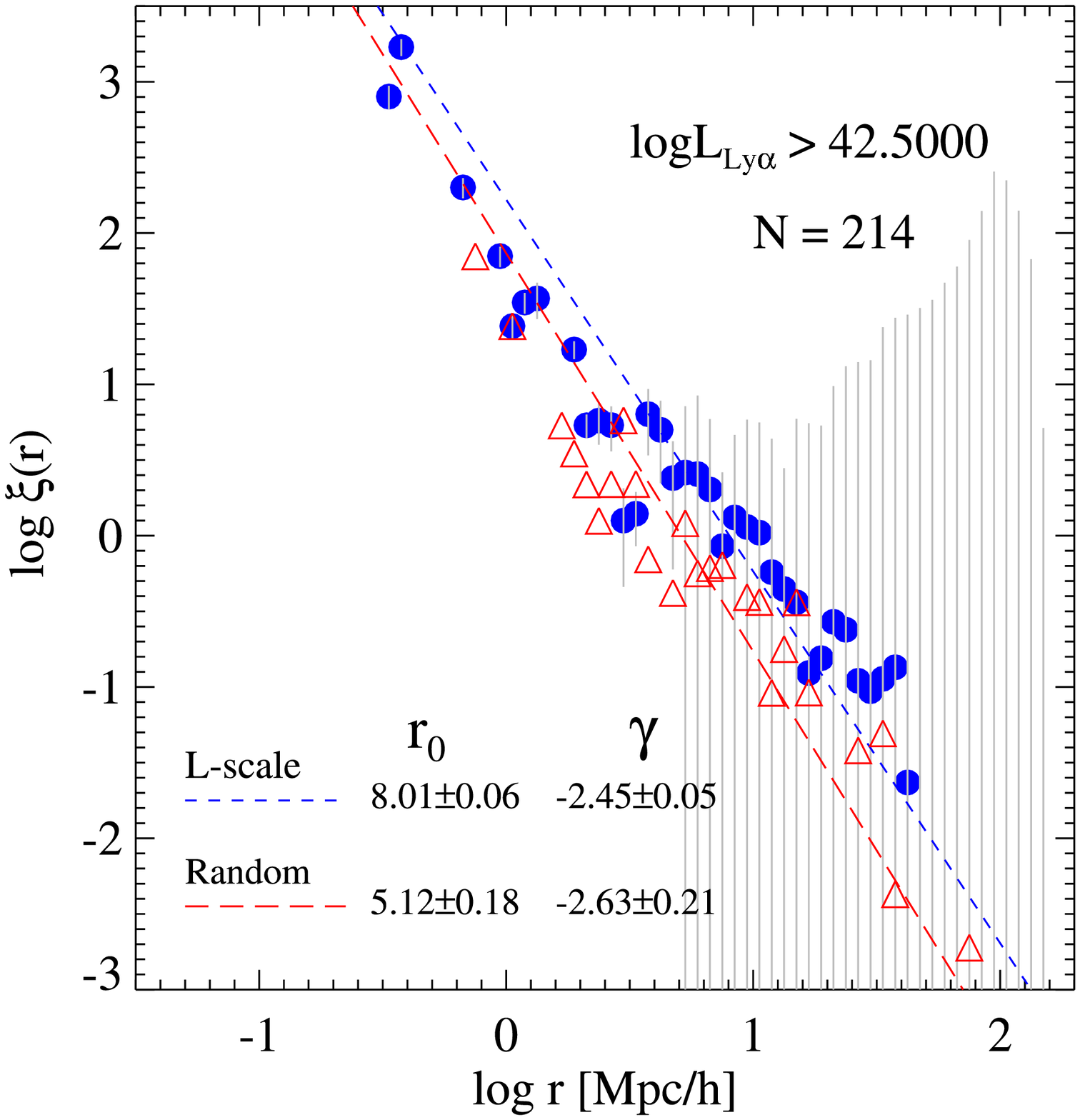,width=2.7in}}
\centerline{\psfig{file=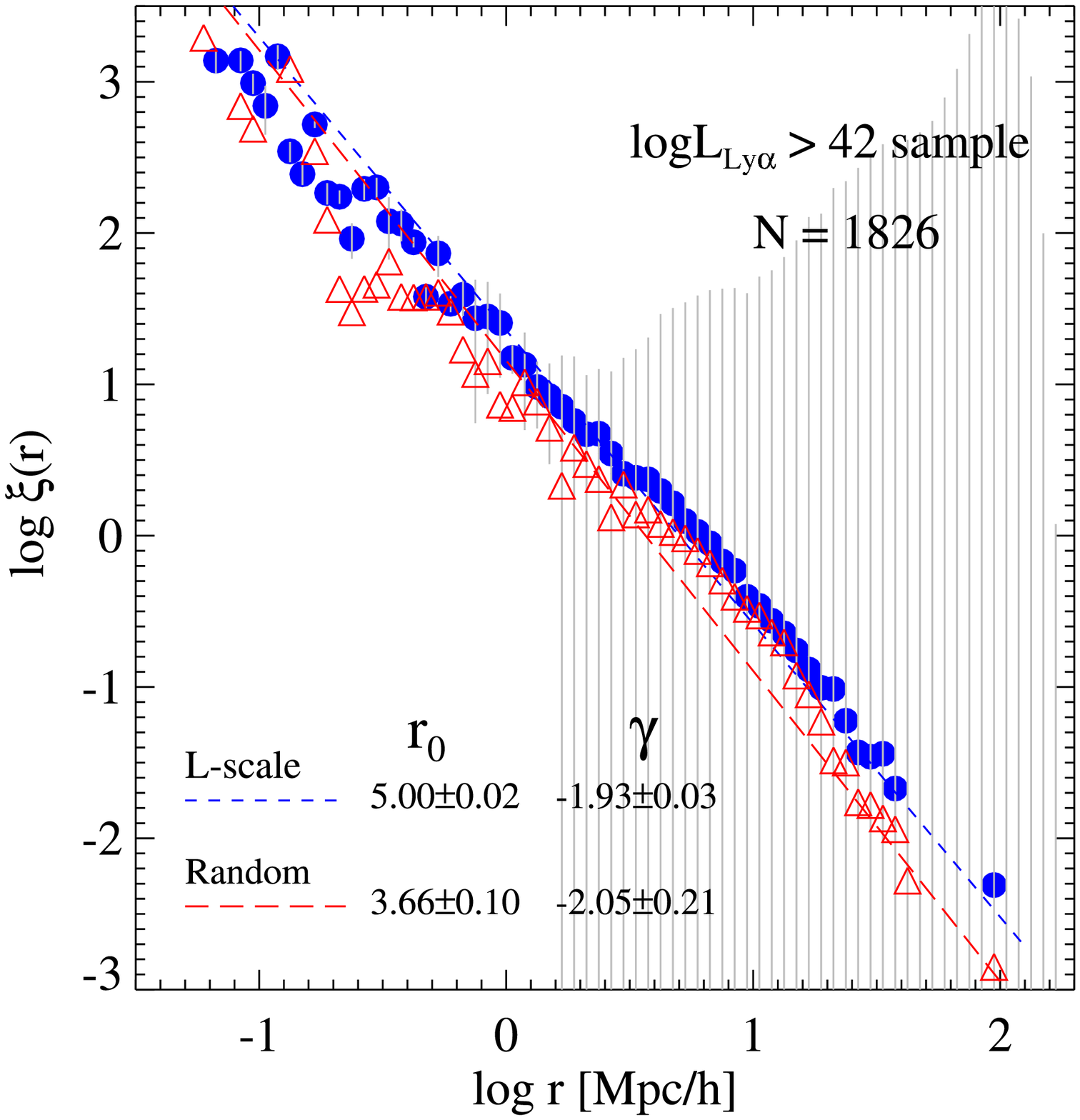,width=2.7in}\psfig{file=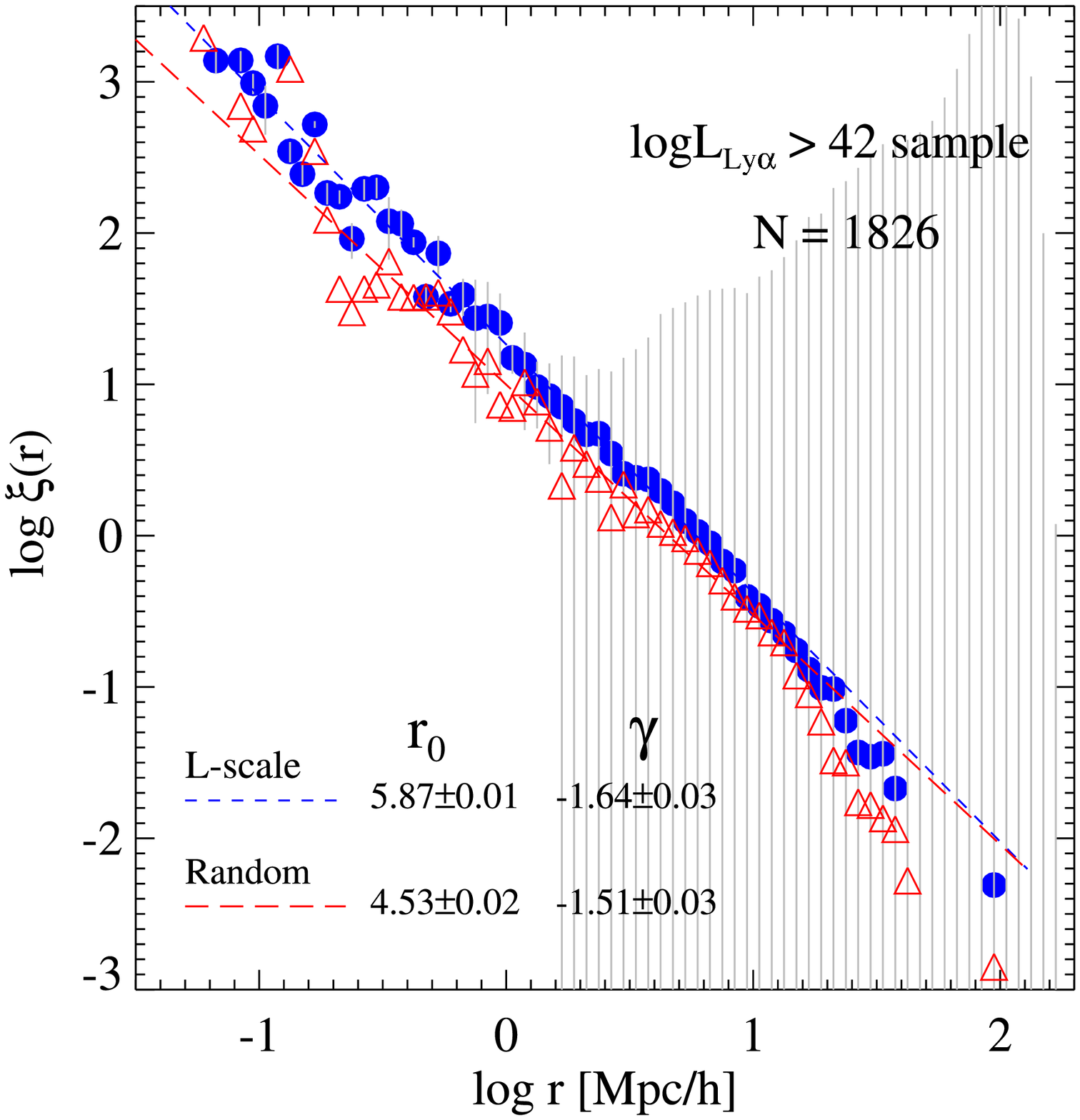,width=2.7in}}
\caption{Correlation function of simulated LAEs in SPH G6 run at $z=6$ 
with different luminosity cuts. In all cases, the `uniform dimming' case 
(solid circles and {\it short-dashed} line) has a longer correlation 
length $r_0$ than the `random sampling' case (open triangles and 
{\it long-dashed} line). The bottom right panel uses the same sample
as the bottom left panel, but the power-law fitting was performed
to the data points limited to the range $-0.5 < \log r < 1.0$. 
The gray error bars are the standard deviations of the distribution 
for each bin based on the calculation of $\xi(r)$ for $2\times 2\times 2=8$ 
smaller subdivided simulation boxes. Therefore they include the 
effect of cosmic variance on scales of comoving 50 $h^{-1}$Mpc, but not for
larger scales.  
}
\label{fig:xi}
\vspace{0.5cm}
\end{figure}

There is a possibility of distinguishing the two scenarios discussed 
in the previous section by looking at the correlation strength of 
the spatial distribution of LAEs. In the `uniform dimming' scenario, 
the sources basically correspond to the brightest galaxies in the 
most massive halos, therefore they are as strongly clustered as the 
bright LBGs. In the `random sampling' scenario, there are two effects
that would weaken the clustering strength compared to the `uniform
dimming' scenario: 1) the number of more massive halos that dominate 
the correlation signal is smaller in the randomly selected sample 
than the parent sample, because the number of lower
luminosity LAEs that are hosted by lower mass halos is larger, and  
2) Poisson noise is introduced by the random sampling. 

This is demonstrated in Figure~\ref{fig:xi}, where the spatial 
correlation function was computed in the two different scenarios 
for samples with different Ly$\alpha$ luminosity limit. 
In all cases, the `uniform dimming' scenario (solid circles and the 
{\it short-dashed} line for the power-law fit) 
has a longer correlation length $r_0$ than the `random sampling' case
(open triangles and the {\it long-dashed} line). There does not 
seem to be a clear trend for the slope of the correlation function. 
Our predictions for the correlation lengths can be directly compared 
with observational data, and the above two scenarios can be tested 
in the near future.


\section{Conclusions \& Discussions}
\label{sec:discussion}

The main conclusions of my talk are the following:

\begin{enumerate}

\item Luminosity functions of simulated galaxies in cosmological 
hydro simulations agree well with those of the observed LBGs at 
$z=3-6$ at the bright-end (Night et al. 2005). At $z=6$, they have 
a steep faint-end slope of $\alpha \sim -2.0$ in the rest-frame 
magnitude range $-19 < M_{1350} <-16$. This magnitude range is even 
beyond the current magnitude limit of rest-frame $M_{1350}\simeq -19$.

\item Cosmological hydro simulations predict that the cosmic 
star formation rate density peaks at $z\ge 5$, unlike most of the 
semi-analytic models of galaxy formation (Nagamine et al. 2004b). 

\item The above two facts suggest that there would be enough ionizing
photons to reionize the Universe at $z=6$ from the Pop II stars in 
ordinary galaxies.

\item The simulations overpredict the LAE LF by a factor of $\sim 10$, 
and two different scenarios (`uniform dimming' and `random sampling') 
can be considered to reconcile this result with the observed LF. 

\item The `uniform dimming' scenario has a stronger spatial correlation 
strength for LAEs than the `random sampling' scenario, and this can be
tested by direct comparison to the observed data. 

\end{enumerate}

In the `uniform dimming' scenario, the LAE sample 
for a given Ly$\alpha$ luminosity cut would directly correspond 
to a sample of bright LBGs that are already observed by the 8 meter 
class telescopes and the Hubble Space Telescope. 
Therefore the `uniform dimming' scenario suggests
a strong link between the LAE and LBG sample, which can also be 
tested by comparing the spatial correlation function of the two 
samples at the same redshift. This observational test can be readily 
performed with the existing Subaru telescope data (Ouchi et al. 2004, 
2005). In the real Universe, it is certainly possible that both 
effects of `uniform dimming' and `random sampling' may be at play.  
We also plan to 
perform direct comparisons of the simulated and the observed angular 
correlation functions of both LAEs and LBGs in a similar spirit to 
the work by Hamana et al. (2004). The physical link between 
LAEs and LBGs can also be tested by a cross-correlation study 
of the two population.


\end{document}